\documentclass[12pt,a4paper]{article}
\usepackage{latexsym}
\usepackage{amsfonts}

\newcommand{\danger}[1]{\textbf{#1}}

\input{epsf}             
\def\epsfcenter#1{{\vcenter{\hbox{\epsfbox{#1}}}}} 

\usepackage[dvips]{graphicx}

\begin{document}

\title{(2+1)-dimensional quantum gravity,
spin networks and asymptotics}
\author{J. Manuel Garc\'\i a-Islas}

\maketitle

\hspace{3.5cm}School of Mathematical Sciences

\hspace{4.0cm} University of Nottingham

\hspace{3.7cm} Nottingham, NG7 2RD, UK

\hspace{2.0cm} email: jm.garcia-islas@maths.nottingham.ac.uk

\vspace{1.3cm}

\danger{Abstract}. A method to evaluate spin networks for
(2+1)-dimensional quantum gravity is given.
We analyse
the evaluation of spin networks for Lorentzian, Euclidean and a
new limiting case of Newtonian quantum gravity.
Particular attention is paid to the tetrahedron and to the study of
its asymptotics.
Moreover, we propose that all this technique can be extended to
spin networks
for quantum gravity in any dimension.

\section{\bf{Introduction}}

In this paper we study the evaluation of spin networks in (2+1) -dimensional
Lorentzian, Euclidean and Newtonian quantum gravity.
The concept of spin networks was first introduced by Penrose in \cite{p},
in order to describe a combinatorial picture of the geometry of
space-time.

The evaluations of spin networks are of great importance in the framework
of Spin Foam models of quantum gravity.
The first example of a Spin Foam model was proposed exactly in (2+1) dimensions.
This is known as the Ponzano-Regge model \cite{pr}, and it is a
model of Euclidean quantum gravity.
This is constructed in a triangulated
(2+1)-dimensional manifold $M$, by assigning an irreducible representation
of $SU(2)$ to each edge of the triangulation.
The corresponding representations are labelled by half integers
$j=0, \frac{1}{2},1,...$. Then to each tetrahedron
we assign a 6j-symbol. We then multiply all of the 6j-symbols assigned to
the tetrahedra of the triangulation of our manifold and then sums over all
the irreduclible representations of our group $SU(2)$.

This is a kind of discretised path integral for Euclidean quantum gravity.
Ponzano and Regge then studied the asymptotics of this Euclidean 6j-symbol.

One question one can ask first is whether a similar Ponzano-Regge
formulation exists for (2+1)-dimensional Lorentzian quantum gravity
and whether we can study the asymptotics of the Lorentzian
tetrahedron network.
This problem was studied first in \cite{d} and in \cite{d2}, by analysing
the asymptotics of a Lorentzian tetrahedron labelled by discrete unitary
representations of the Lorentz group.
Then in \cite{f1} a state sum model for (2+1)-dimensional Lorentzian quantum
gravity was proposed.
Later the asymptotics of
the square of the
Lorentzian 6j-symbol was studied in \cite{f2}. One of the new problems we study in
this paper is the asymptotics of the usual Lorentzian tetrahedron network
labelled by continuous unitary irreducible representations of
the connected Lorentz group $SO_{0}(2,1)$.
Inspired by \cite{bc} we study, in a similar fashion,
Lorentzian (2+1)-dimensional quantum gravity. Particularly we pay attention
to the tetrahedron which is of great importance in the evaluation of transition
amplitudes and of the partition function.
There are different kinds of tetrahedra in Lorentzian geometry according to
whether their corresponding edge vectors are timelike, spacelike or null.
When dealing with spin foam models for quantum gravity the
language of representation theory comes into play and the Lorentzian tetrahedra
spin networks may have
edge vectors labelled by any mixture
of discrete, or continuous representations of the
connected Lorentz group $SO_{0}(2,1)$.
When considering the continuous principal unitary representations only
we can
attempt to construct a spin foam model for (2+1)-dimensional Lorentzian
quantum gravity by using the dual
complex $\mathcal{J}_{\Delta}$ to a
triangulation $\Delta$ of our 3-dimensional manifold.

In order to construct the model we label each face of the complex
$\mathcal{J}_{\Delta}$ by a
principal unitary irreducible representation of the connected Lorentz group
$SO_{0}(2,1)$, or equivalently if we think of the
triangulation, we label each edge instead.
The state sum model is then given by

\begin{equation}
\mathcal{Z}(M)= \int_{0}^{\infty} d\rho_{f} \prod_{f} A(f) \prod_{e} A(e)
\prod_{v} A(v)
\end{equation}
where the integration is carried over the labels of all internal faces of the
dual complex and
$A(f), A(e), A(v)$ are the face, edge, and vertex amplitudes
respectively.

These amplitudes are given by the evaluation of spin networks such as

\[ A(f)= \epsfcenter{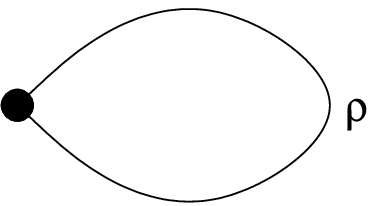} \]

\[ A(e)= \frac{1}{\epsfcenter{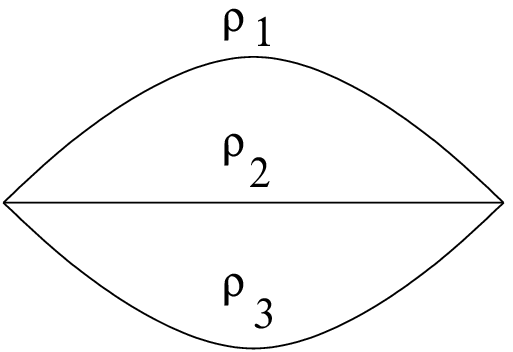}} \]

\[ A(v)= \epsfcenter{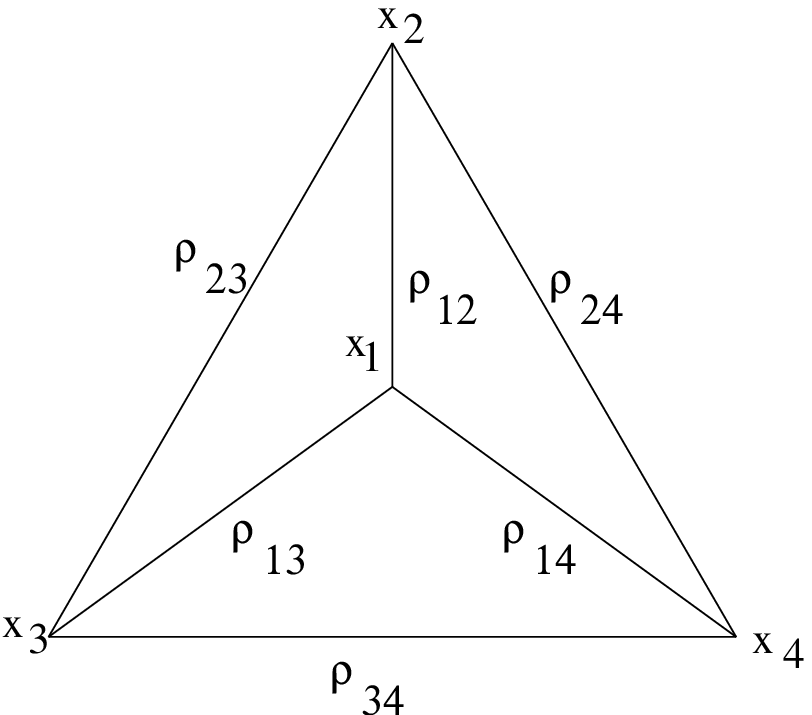} \]

The definition for evaluating spin networks for this (2+1)-dimensional
Lorenztian model is one of the problems which concerns us in this paper.
In particular it is the evaluation of the tetrahedron network(vertex amplitude),
that we concentrate on.

The way to evaluate our spin networks is based on a similar idea
first described in \cite{b1} for the evaluation of relativistic spin networks.
This method was then applied to the study of the 4-simplex in \cite{bw}.
The idea was interpreted to the case of simple spin networks in \cite{fk}
by thinking of them as Feynman integrals
over an internal space. This idea
is given by using propagators(or kernels) which are defined in the internal space.
Then with every edge of the spin network we associate a propagator, and then we
take a product of all these kernels and integrate over a copy of the internal space
for each vertex.

Particularly for our Lorentzian case model the above recipe, applied here for the first
time, translates as follows:
The first ingredient we need for
our description is a kernel $K_{\rho}(x,y)$
where $\rho$ is a real number which stands for a unitary principal irreducible
representation of $SO_{0}(2,1)$ and $x$ and $y$ are two points in the hyperbolic
plane $H^{2}$. The latter can be seen as the upper sheet of a timelike
two-sheeted hyperboloid in 3-dimensional Minkowski space-time.

Once we have this kernel $K_{\rho}(x,y)$, in particular the evaluation of the tetrahedron
will be given by the following integral

\begin{eqnarray}
\int_{H^{2} \times H^{2} \times H^{2}} dx_{2} dx_{3} dx_{4}
K_{\rho_{12}}(x_{1},x_{2}) K_{\rho_{13}}(x_{1},x_{3})
K_{\rho_{14}}(x_{1},x_{4}) \nonumber\\
\qquad \qquad \times K_{\rho_{23}}(x_{2},x_{3})
K_{\rho_{24}}(x_{2},x_{4}) K_{\rho_{34}}(x_{3},x_{4})
\end{eqnarray}
where $x_{1}, x_{2}, x_{3}, x_{4} \in H^{2}$ and our integral is computed over
3 copies of the hyperbolic plane\footnote{Originally we would
have an integral over 4 copies of $H^{2}$, but regularisation requires
that one of these integrals is dropped}.

One of the problems is to know whether our tetrahedron evaluation converges.
Once this is true
our next step is to study the asymptotics of the
above integral.

This way of evaluating spin networks gives us
an attempt to define the computation of transition amplitudes between spin
networks by using the spin foam model description of formula (1).
This gives us the analogue of
the previous known Euclidean Ponzano-Regge model \cite{pr}.

Then it is immediately logical to ask whether we can apply the same technique
to the Euclidean model and to the asymptotics of the tetrahedron network.

This is then our next step, to study
in a similar fashion the case of Euclidean quantum gravity whose group is given
by $SO(3)$.
The irreducible
representations of $SO(3)$ are given by positive integers.
Then again, if we
think of the dual complex of our triangulation, we can define a spin foam
formula for Euclidean quantum gravity given by

\begin{equation}
Z(M)= \sum_{j=0}^{\infty} \prod_{f} A(f) \prod_{e} A(e)
\prod_{v} A(v)
\end{equation}
This is the analogue of our integral formula (1), and moreover, it is very similar
to the Ponzano-Regge formula. The important difference is to notice that
in the Ponzano-Regge model we have the group $SU(2)$ which
is the double covering of $SO(3)$.
Then we go on to calculating the amplitudes for Euclidean spin networks by
using the described recipe. This case has been studied before
in \cite{fk}. However our treatment
completes the whole story, as we include all the details of the calculation of the
asymptotics of the tetrahedron network by using the stationary phase method and moreover, we
mention the case of degenerate configurations.

The evaluation goes as follows:
considering a kernel $k_{\ell}(x,y)$, where $\ell$ denotes an irreducible
representation of $SO(3)$.  For this case $x,y$ are points on
the two dimensional sphere $S^{2}$.
Our tetrahedron network will be given by the integral

\begin{eqnarray}
\int_{(S^{2})^{4}} dx_{1} dx_{2} dx_{3} dx_{4}
K_{\ell_{12}}(x_{1},x_{2}) K_{\ell_{13}}(x_{1},x_{3})
K_{\ell_{14}}(x_{1},x_{4}) \nonumber\\
\qquad \qquad \times K_{\ell_{23}}(x_{2},x_{3})
K_{\ell_{24}}(x_{2},x_{4}) K_{\ell_{34}}(x_{3},x_{4})
\end{eqnarray}
In this case we do have an integral over 4 copies of $S^{2}$ and no regularisation
is required, as $S^{2}$ is compact.

We complete our picture by studying a completely new case, at least in
the (2+1)-dimensional quantum gravity case, which
we call Newtonian quantum gravity.
This case deals with $ISO(2)$ spin networks.
The construction of such kind of quantum gravity is analogous to the Lorentzian
case, as principal unitary irreducible representations of $ISO(2)$ are also labelled by
positive real numbers $\mathbb{R}^{+}$. The evaluation of the tetrahedron
network for this case is given by a similar integral to integral (2), with the
only difference that for the Newtonian case, we integrate over three copies
of the flat plane $\mathbb{R}^{2}$.

The idea is that all these three
cases are related to each-other by a limit process, and then we have a whole
picture of a unified theory. This unified theory idea was also proposed in
\cite{bce} for the (3+1)-dimensional quantum gravity case.

The main questions we ask then are: What do the kernels look like?
What are the evaluations of the tetrahedron networks in each case?

We divide this paper in the following sections. In section 2 we study
the case of (2+1)-dimensional Lorentzian quantum gravity
spin networks. The computation of the spin networks with the
help of an internal space is completely new. The whole idea
and techniques were discovered by following a similar method to the one used
for the (3+1)-dimensional Lorentzian Barrett-Crane model.

In section 3 we study with the same spirit the model
of Euclidean spin networks.

In section 4 we study a new kind of model
which we call Newtonian spin networks. The kind of spin netwoks for this
model were studied for the case of 4-dimensions in \cite{bce}. However
for (2+1)-dimensions is completely new as well as the name we give to it.
The explanation of why we call them Newtonian is also new.

Finally in section 5 we conclude by proposing a way to study spin networks for quantum
gravity in any dimension following the spirit of the paper and which
generalises our (2+1)-dimensional case. Furthermore for (3+1)-dimensions
we just have again the well known Barrett-Crane evaluations.

\section{\bf{Lorentzian spin networks}}

Equation (1) gives a way to define a partition function for the
(2+1)-dimensional Lorentzian quantum gravity as a spin foam model, for
the case in which we are only considering the principal unitary representations
of the group $SO_{0}(2,1)$.

In this section we define the evaluation of spin networks and
construct the whole picture behind
the paper. It is in the same spirit of this section that all of the future
sections follow.

\subsection{\bf{$SO_{0}(2,1)$ and its principal unitary representations}}

Recall 3-dimensional Minkowski space-time to be
$\mathbb{R}^{3}$ with its Lorentzian
metric given by

\[ [x,y]= x_{0}y_{0}-x_{1}y_{1}-x_{2}y_{2} \]
where $x=(x_{0},x_{1},x_{2}), y=(y_{0},y_{1},y_{2}) \in \mathbb{R}^{3}$
and $x_{0}$ and $y_{0}$
are the time components.

Consider the following subsets of Minkowski space:

1) The Cone $C$: This is the set of points of Minkowski space which satisfy
$[x,x]=0$, that is, the set of vectors which are at a distance zero
from the origin. These are called the
lightlike vectors as this set describes particles which travel at the speed
of light.

The upper cone $C^{+}$: This is the set of points in $C$ for which the time
component $x_{0}>0$.

2) The two-dimensional hyperbolic space $H^{2}$: This is the set of points for
which $[x,x]=1$ and $x_{0}>0$.

Consider $SO(2,1)$, the group of unimodular linear transformations of
Minkowski space-time which preserves the Lorentzian metric. The subgroup of
$SO(2,1)$ which preserves $C^{+}$ is denoted $SO_{0}(2,1)$. This group is
connected and locally compact.

In fact the action of $SO_{0}(2,1)$ in $\mathbb{R}^{3}$ gives an action
of $SO_{0}(2,1)$ in the hyperbolic space $H^{2}$, and the subgroup of
$SO_{0}(2,1)$ which leaves the vector $(1,0,0)$ fixed, is isomorphic to
$SO(2)$. It is a maximal compact subgroup and therefore $H^{2}$ can equivalently
be seen as a homogeneous space given by the quotient $SO_{0}(2,1)/SO(2)$.

The representations of $SO_{0}(2,1)$ are realised in the following way.
First of all, we introduce spherical coordinates on
the hyperbolic space $H^{2}$. That is, for $x=(x_{0},x_{1},x_{2}) \in H^{2}$
its coordinates are given by

\begin{eqnarray}
x_{0}= \cosh r  \nonumber \\
x_{1}= \sinh r \sin \theta \nonumber \\
x_{2}= \sinh r \cos \theta
\end{eqnarray}
In these coordinates the Lorentzian wave operator given by

\begin{equation}
\Delta= - \frac{\partial^{2}}{\partial x_{0}^{2}}+
\frac{\partial^{2}}{\partial x_{1}^{2}} + \frac{\partial^{2}}{\partial x_{2}^{2}}
\end{equation}
when restricted to the hyperbolic space $H^{2}$ is given by

\begin{equation}
\Delta_{+} = (\sinh r)^{-1} \frac{\partial}{\partial r}
\sinh r \frac{\partial}{\partial r} + (\sinh r)^{-2}
\frac{\partial^{2}}{\partial \theta^{2}}
\end{equation}
Consider the space of functions inside $C^{+}$. The group $SO_{0}(2,1)$
acts on this space and the operator of equation (7) commutes with this action.
Then if there is a function $f$ for which $\Delta f(x)=0$ then
$\Delta f(g^{-1} x)=0$, for $g \in S_{0}(2,1)$. A function which satisfies
$\Delta f(x)=0$ is called harmonic. Consider the space of harmonic functions
defined inside $C^{+}$ of degree $\sigma \in \mathbb{C}$. The equality
$T^{\sigma}(g)f(x)=f(g^{-1}x)$ defines a representation of $SO_{0}(2,1)$
in the space of harmonic functions.
There are the
ones called $principal$ $representations$ which are the ones
we are interested in in this work. These representations are labelled as
$(i \rho - \frac{1}{2})$, where $\rho$ is a real parameter.
Its importance for the physics of this work is that tetrahedra embedded
in 3-dimensional Minkowski space for which edges are
labelled by principal unitary  representations are thought of as
corresponding to  spacelike tetrahedra, that is all their edge vectors are
spacelike.

\subsection{\bf{The kernel and its asymptotics}}

As in \cite{bc} we approach to the calculation of the kernel which is the main
ingredient for the evaluation of the
spin networks, and most important to our case of the evaluation of the
tetrahedron.

Given a function $h \in L^{2}(H^{2})$ we have the $\rho$ irreducible component

\begin{equation}
 f_{\rho}(\xi)= \int_{H^{2}} h(x) [x, \xi]^{i \rho - \frac{1}{2}} dx
\end{equation}
where $\xi$ is a lightlike vector, that is $\xi \in C^{+}$.
According to \cite{vk}, the inverse transform is given by an integral of the
type

\begin{equation}
h(x)= \int_{0}^{\infty} d\mu_{\rho} \int_{S^{1}}
f_{\rho}(\xi) [x, \xi]^{-\frac{1}{2}- i \rho} d \xi
\end{equation}
where $S^{1}$ is a contour on $C^{+}$ intersecting every generatrix of the cone
at one point. More specifically for our purpose it is taken to be the
unit circle $S^{1}$.
Composing these two transforms without integrating over $\rho$ gives the
projection operator onto the $\rho$
irreducible component $h_{\rho}$ given by

\begin{equation}
h_{\rho}(x)= \frac{1}{4i \pi} \int_{H^{2}} \int_{S^{1}}
[x, \xi]^{- \frac{1}{2}-i \rho} [y, \xi]^{- \frac{1}{2}+i \rho}
h(y) d \xi dy
\end{equation}
where we consider the kernel function

\begin{equation}
K_{\rho}(x,y)= \int_{S^{1}} [x, \xi]^{- \frac{1}{2}-i \rho}
[y, \xi]^{- \frac{1}{2}+i \rho} d \xi
\end{equation}
where $x,y \in H^{2}$,
and then

\begin{equation}
h_{\rho}(x)= \frac{1}{4i \pi} \int_{H^{2}} K_{\rho}(x,y) h(y) dy
\end{equation}
We now consider the kernel and express it in a spherical coordinate system. For
this to be done, we shift the points $x,y \in H^{2}$ (which always can be done)
to the following points

\[ x=(\cosh r, 0, \sinh r) \hspace{1.5cm} y=(1,0,0) \]
by abuse of notation we still call them $x$ and $y$. Moreover
$\cosh r= [x,y]$ is the hyperbolic distance between $x,y \in H^{2}$.
The point $\xi \in S^{1}$ has coordinates $(1, \sin \theta, \cos \theta )$.
Hence we write

\begin{equation}
K_{\rho}(x,y)= \frac{1}{2 \pi}\int_{0}^{2 \pi}
( \cosh r - \cos(a) \sinh r)^{- \frac{1}{2}-i \rho} da
\end{equation}
It turns out that our kernel is indeed
the well known zonal spherical functions for representations of the group
$SO_{0}(2,1)$, known also as Legendre functions $\mathfrak{B}_{\tau}(\cosh r)$
where $\tau = -1/2 - i \rho$.
Some of its properties are described as follows.

This kernel is an eigenfunction of the Laplace operator on $H^{2}$.
That is, $\Delta_{+} K_{\rho}(x,y)= -(1/4 + \rho^{2})K_{\rho}(x,y)$. Because of
this property, the
function $K_{\rho}(x,y)$  is called
a spherical harmonic.

Now, given a spin network with edges labelled by real numbers $\mathbb{R}^{+}$
corresponding to principal unitary irreducible representations, we evaluate
it as follows. To each vertex we associate a variable $x \in H^{2}$. To
each edge labelled by $\rho$, we associate the kernel $K_{\rho}(x,y)$ where
the variables $x$ and $y$ are the ones associated to its correspondent
vertices.

The evaluation is then given by the integral

\begin{equation}
 \int_{(H^{2})^{n-1}} \prod_{i<j}
K_{\rho_{ij}}(x_{i}, x_{j}) dx_{2}...dx_{n}
\end{equation}
where, as in \cite{bc}, we have regularized the evaluation by removing the
integration over one of the variables in $H^{+}$ at one of the vertices.
This variable is then fixed.

From this definition we can attempt to evaluate some spin networks. It follows
easily for example that the evaluation for our face amplitude

\[ A(f)= \epsfcenter{lor3.eps} \]
gives

\[ K_{\rho}(x,x)= 1 \]
Some other examples such as our edge and vertex amplitudes result in
integrals which are difficult to evaluate directly. So we overcome this
problem by considering the asymptotics of one of these evaluations, which is
of our main concern. We then study the asymptotics of the tetrahedron network.

Given our tetrahedron network labelled by principal
unitary irreducible
representations $\rho_{12}, \rho_{13}, \rho_{14}, \rho_{23}, \rho_{24},
\rho_{34}$ as in figure 1.

\begin{figure}[h]
\begin{center}
\includegraphics[width=0.3\textwidth]{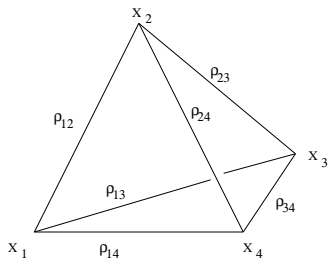}
\caption{Tetrahedron network}
\end{center}
\end{figure}

Naively we evaluate it as the following integral

\begin{eqnarray}
T_{6}= \int_{H^{2} \times H^{2} \times H^{2}} dx_{2} dx_{3} dx_{4}
K_{\rho_{12}}(x_{1},x_{2}) K_{\rho_{13}}(x_{1},x_{3})
K_{\rho_{14}}(x_{1},x_{4}) \nonumber\\
\qquad \qquad \times K_{\rho_{23}}(x_{2},x_{3})
K_{\rho_{24}}(x_{2},x_{4}) K_{\rho_{34}}(x_{3},x_{4})
\end{eqnarray}
Note that this is an integral involving other integrals, that is, the kernels
which are also given by integrals.
We therefore have to know whether our tetrahedron evaluation converges. We do not have
a way to prove that our tetrahedron integral is finite but we give some evidence
about it. First of all the problem is very similar to the 4-dimensional Barrett-Crane
model of evaluating the 10j symbol, and this is being proved to be finite.

We can prove the convergence of the theta symbol. This latter is given as follows:
We have that our Lorentzian kernel when the distance $r$ between the points
$x$ and $y$ goes to infinity and $\rho \neq 0$, \cite{t},
is asymptotic to

\begin{equation}
K_{\rho}(x,y) \sim A \frac{1}{(2 \cosh r)^{1/2}}
\end{equation}
where $A$ is a factor which oscillates.

Then for very large $r$ we have that our kernel is asymptotic to
$\frac{1}{\sqrt{2}}e^{-r/2}$. With this asymptotics
our theta symbol evaluation is then given by

\begin{equation}
\epsfcenter{lor2.eps} \sim 2 \pi \int_{}^{\infty} \frac{1}{(2 e^{r})^{3/2}} e^{r} dr
\end{equation}
which is finite.

We then have a finite value of our theta symbol and this makes
the convergence of our tetrahedron
more promising. More evidence is given is section 4 when dealing with our Newtonian
quantum gravity model;
the Lorentzian and Newtonian models are
related by a limit as explained in section 4.

Supposing then that our tetrahedron evaluation converges, we study the
asymptotics of this integral.

As a warm-up example let us compute
the asymptotics of our kernel $K_{\rho}(x,y)$ first.

Recall that the general method of stationary phase states that an
n-dimensional integral of
the form

\begin{equation}
I(x)= \int g(x) e^{i \alpha f(x)} dx
\end{equation}
has an asymptotic expansion when $\alpha \rightarrow \infty$ given by

\begin{equation}
I(x) \sim \biggl( \frac{2 \pi}{\alpha} \biggr)^{n/2}
\sum_{x \mid df(x)=0} g(x) e^{i \alpha f(x)}
\frac{e^{i \pi sgn(H)/4}}{\mid det H(x) \mid^{1/2}} + O(\alpha^{-n/2 -1})
\end{equation}
where $H(x)$ is the Hessian matrix for $f(x)$, and $sgn(H)$ is its signature,
i.e the number of positive eigenvalues minus the number of negative eigenvalues.

Now observe that for the case of our kernel we have
$g(x)=( \cosh r- \cos(x) \sinh r)^{- \frac{1}{2}}$ and
$e^{ik \phi(x)}= e^{i \rho (- \ln ( \cosh r - \cos(x) \sinh r))}$, so that
$\phi(x)=- \ln ( \cosh r - \cos(x) \sinh r)$. Then

\begin{equation}
\phi^{'}(x)= - \frac{ \sinh r \sin(x)}{ \cosh r - \cos(x) \sinh r}
\end{equation}
and so the stationary points of $\phi$ are given by $x=0$ and $x= \pi$ if
$r \neq 0$.

Then following the stationary phase formula, the asymptotics of $K_{\rho}(x,y)$
as $\rho \rightarrow \infty$ is given up to a $1/2 \pi$ factor by

\begin{equation}
K_{\rho}(x,y) \sim \biggl( \frac{2 \pi}{\rho \sinh r} \biggr)^{\frac{1}{2}}
\biggl[ \frac{e^{-i \pi/4}}{( \cosh r - \sinh r)^{i \rho}} +
\frac{e^{i \pi/4}}{( \cosh r + \sinh r)^{i \rho}} \biggr]
\end{equation}
Using trigonometric identities it is easy to see that our above
formula reduces to

\begin{equation}
K_{\rho}(x,y) \sim \biggl( \frac{8 \pi}{\rho \sinh r} \biggr)^{\frac{1}{2}}
\biggl[ e^{i( \rho r - \pi/4)} + e^{-i( \rho r - \pi/4)}
\biggr]
\end{equation}
which can also be seen as

\begin{equation}
 K_{\rho}(x,y) \sim \biggl( \frac{8 \pi}{\rho \sinh r} \biggr)^{\frac{1}{2}}
\cos( \rho r - \pi/4)
\end{equation}

\subsection{\bf{The tetrahedron network}}

On the assumption that our tetrahedron network is finite,
we now proceed to compute the asymptotics of formula (2). For that purpose
we replace the labels $\rho_{ij}$ by $\alpha \rho_{ij}$ to study the
asymptotics when $\alpha \rightarrow \infty$.
We then have

\begin{equation}
T_{6}= \int_{({H^{2}})^{3}} \prod_{i<j}
\int_{[0, 2 \pi]^6}
( \cosh r_{ij} - \cos(\theta_{ij}) \sinh r_{ij})^{- \frac{1}{2}+i \alpha \rho_{ij}}
d \theta_{ij} dx_{j}
\end{equation}
which we now write as

\begin{eqnarray}
T_{6}= \int_{({H^{2}})^{3}} \prod_{i<j}
\int_{[0, 2 \pi]^6}
( \cosh r_{ij} - \cos(\theta_{ij}) \sinh r_{ij})^{- \frac{1}{2}} \nonumber\\
\qquad \qquad \qquad \times \biggl[ e^{i \alpha \sum_{i<j} \rho_{ij}
\ln ( \cosh r_{ij} - \cos(\theta_{ij}) \sinh r_{ij})}
\biggr] d \theta_{ij} dx_{j}
\end{eqnarray}
The exponent $S= \sum_{i<j} \rho_{ij}
\ln ( \cosh r_{ij} - \cos(\theta_{ij}) \sinh r_{ij})$ in the
equation above is called the action from which we have to compute the stationary
points in order to obtain the asymptotics of our tetrahedron network.

Solutions for which our action is stationary should be classified as either
degenerate ones or non-degenerate. Degenerate solutions
correspond to a degenerate tetrahedron, that is, one for which any or all of
its vertices coincide,
$x_{i}=x_{j}$ for any $i \neq j$, or equivalently when some or all $r_{ij}=0$.

Non-degenerate solutions correspond to tetrahedra for which $r_{ij} \neq 0$
for all $i \neq j$.

We then start our study of all these possible solutions and follow
a procedure, similar to the one used in \cite{bs}

\subsubsection{Non-degenerate solutions}

We have in this case that all of the points
$x_{i} \in H^{2}$ are different, that is $r_{ij} \neq 0$ for all $i,j$.

We start varying the action

$S= \sum_{i<j} \rho_{ij} \ln ( \cosh r_{ij} - \cos(\theta_{ij}) \sinh r_{ij})$
with respect to $\theta_{ij}$ and
$x_{j}$. We use a Lagrange multiplier $\lambda_{j}$ for the constraint we have
$[x_{j},x_{j}]=1$.

A first set of equations are given by varying our action with respect
to $\theta_{ij}$, that is

\begin{equation}
\frac{\partial S}{\partial \theta_{ij}} = 0
\end{equation}
from which it follows

\begin{equation}
\frac{\rho_{ij} \sin(\theta_{ij}) \sinh r_{ij}}
{\cosh r_{ij} - \cos(\theta_{ij}) \sinh r_{ij}} = 0
\end{equation}
Then the above equation, for the case for which $r_{ij} \neq 0$ for all
$i \neq j$,
is satisfied when $\theta_{ij}= 0$ or $\theta_{ij}= \pi$

To keep track of the possible choices of $\theta_{ij}=0$ or $\theta_{ij}= \pi$
we introduce a variable $\epsilon_{ij}=-1$ if $\theta_{ij}=0$ or
$\epsilon_{ij}=1$ if $\theta_{ij}=\pi$.

Our action can then be written $S=\sum_{i<j} \epsilon_{ij} \rho_{ij} r_{ij}$
so that a second set of equations are given by varying our action with
respect to
$x_{2}, x_{3}, x_{4}$ that are the variables of our integrand.
We take care of our constraint mentioned before, that is

\begin{equation}
\frac{\partial S}{\partial x_{j}} = \lambda_{j} x_{j}
\end{equation}
where $j=2,..,4$.

From this it is easy to see that we arrive at the following equations

\begin{eqnarray}
\frac{\epsilon_{12} \rho_{12} x_{1}}{\sinh r_{12}} +
\frac{\epsilon_{23} \rho_{23} x_{3}}{\sinh r_{23}} +
\frac{\epsilon_{24} \rho_{24} x_{4}}{\sinh r_{24}}=
\lambda_{2} x_{2} \nonumber\\
\frac{\epsilon_{13} \rho_{13} x_{1}}{\sinh r_{13}} +
\frac{\epsilon_{23} \rho_{23} x_{2}}{\sinh r_{23}} +
\frac{\epsilon_{34} \rho_{34} x_{4}}{\sinh r_{34}}=
\lambda_{3} x_{3} \nonumber\\
\frac{\epsilon_{14} \rho_{14} x_{1}}{\sinh r_{14}} +
\frac{\epsilon_{24} \rho_{24} x_{2}}{\sinh r_{24}} +
\frac{\epsilon_{34} \rho_{34} x_{3}}{\sinh r_{34}}=
\lambda_{4} x_{4}
\end{eqnarray}
where we recall that all of our vectors $x_{i}$ are timelike and future
directed, and our $\epsilon$ are the result of the possible election of
the $\theta$ stationary points at $0$ or $\pi$.

We now take the wedge product with $x_{i}$ to the $i$ equation($i=2,..4$),
\footnote{Recall that, $(x_{j} \wedge x_{i})^{a} = \eta^{ae}
\epsilon_{ebc} x_{j}^{b} x_{i}^{c}$, where $\epsilon_{abc}$ is the totally
antisymmetric tensor with $\epsilon_{123}=1$ and $\eta_{ab}= \eta^{ab}$ is
the $-++$ Lorentzian metric.}
where
we take a unit vector $v_{ij}$ in the direction $x_{i} \wedge x_{j}$.
So we have the following set of equations,

\begin{eqnarray}
- \epsilon_{12} \rho_{12} v_{12} +
\epsilon_{23} \rho_{23} v_{23} +
\epsilon_{24} \rho_{24} v_{24}= 0 \nonumber\\
- \epsilon_{13} \rho_{13} v_{13} -
\epsilon_{23} \rho_{23} v_{23} +
\epsilon_{34} \rho_{34} v_{34}= 0 \nonumber\\
- \epsilon_{14} \rho_{14} v_{14} -
\epsilon_{24} \rho_{24} v_{24} -
\epsilon_{34} \rho_{34} v_{34}= 0
\end{eqnarray}
If we sum the three equations above, we get a fourth equation given by

\begin{equation}
\epsilon_{12} \rho_{12} v_{12} +
\epsilon_{13} \rho_{13} v_{13} +
\epsilon_{14} \rho_{14} v_{14}= 0
\end{equation}
If we call $V_{ij}= \epsilon_{ij} \rho_{ij}v_{ij}$, then
the equations (28) and (29), imply that the $V_{ij}$ are the edge vectors of
a tetrahedron with edge lengths $\rho_{ij}$.
Each equation describes a face of a
tetrahedron. We have that the normals to each of these faces are timelike
vectors,  therefore our tetrahedron is spacelike which implies
then that
the usual triangle inequalities should be satisfied for the lengths of its
edges.
Each of our timelike $x_{i}$ normal vectors could be pointing inwards or
outwards, but always future directed. Let $n_{i}$ be the outward normal
pointing vectors to the faces of the tetrahedron. Then we have that
$x_{i}= a_{i} n_{i}$, where $a_{i}= \pm 1$. Recall that our point
$x_{1}$ is fixed at $(1,0,0)$.

All the possible configurations of a spacelike tetrahedron
are divided in two possible configurations only:

1) One normal vector, say $n_{1}$, is future directed, while the remaining
three normal vectors $x_{i}$ for $i=2,3,4$ are past directed.

2) Two normal vectors are future directed and two normal vectors are
past directed.

We could also have one normal vector, say $n_{1}$, is past directed, while
the remaining three normal vectors $x_{i}$ for $i=2,3,4$ are future directed.
But this latter configuration is equivalent to the first one as we need only
to change all the signs.

We then have that the possible values of the $\epsilon$'s are:

For configuration
1), we have that $x_{1}=n_{1}, x_{2}= -n_{2}, x_{3}= -n_{3}, x_{4}= -n_{4}$,
therefore $\epsilon_{12}= -1, \epsilon_{13}= -1, \epsilon_{14}= -1,
\epsilon_{23}= 1, \epsilon_{24}= 1, \epsilon_{34}= 1$.

For configuration 2), we have that
$x_{1}= n_{1}, x_{2}= n_{2}, x_{3}= -n_{3}, x_{4}= -n_{4}$, therefore
$\epsilon_{12}= 1, \epsilon_{13}= -1, \epsilon_{14}= -1,
\epsilon_{23}= -1, \epsilon_{24}= -1, \epsilon_{34}= 1$.

Other equivalent configurations are given by swapping all the signs of the
above configurations.

The Hessian is a $11 \times 11$ matrix with components
$\frac{\partial^{2}S}{\partial x_{i} \partial x_{j}},
\frac{\partial^{2}S}{\partial x_{i} \partial \theta_{ij}},
\frac{\partial^{2}S}{\partial \theta_{ij} \partial \theta_{ij}}$.
When evaluating the Hessian at the stationary points, many of the entries
will have zeros and it is seen that its determinant equals
the determinant of the $5 \times 5$ matrix with
components $\frac{\partial^{2}S}{\partial x_{i} \partial x_{j}}$
evaluated at the stationary points, times some factors.
The remaining factors are given by
$(\rho_{12}...\rho_{34})/
(\prod_{i<j} \cosh r_{ij} - \cos(\theta_{ij}) \sinh r_{ij})^{- \frac{1}{2}})$
which denominator cancels in the evaluation of the asymptotic expansion.

Our integral is of dimension 11 which then implies that there is an
extra factor of $(2 \pi/ \alpha)^{11/2}$

The contribution we will get will be an oscillating function as

\begin{equation}
T_{6} \sim A \hspace{0.2cm}
\cos \biggl( \sum_{k<l} \rho_{ij} r_{ij}+
k \frac{\pi}{4} \biggr)
\end{equation}

There is a factor $A$ containing information about the
determinant of our Hessian at each stationary point.

The information contained in the factor $A$ is related to the volume
of the tetrahedron. This specific calculation can be done numerically for
some examples, but a rigorous proof must be done in a different way.

\subsubsection{Degenerate solutions}

We now proceed to the study of the contribution which comes from degenerate
solutions. Let us deal with the case in which all our vectors $x$'s are
parallel.
We have noted before that our kernel was not very promising when dealing with
this case. However, we change the way our kernel looks so that we may deal
with our degenerate case.
For this purpose we introduce new coordinates as follows:
given two points $x_{i}$ and $x_{j}$ in $H^{2}$ there is a geodesic $\gamma(t)$
which joins the points $x_{i}$ and $x_{j}$. We define the vector $\xi_{ij}$ to
be the initial velocity of our geodesic. Then we have
$\mid \xi_{ij} \mid =r_{ij}$ is the distance between $x_{i}$ and $x_{j}$.
Recall our kernel

\begin{equation}
K_{\rho}(x_{i},x_{j})= \int_{0}^{2 \pi}
( \cosh r - \cos(a) \sinh r)^{- \frac{1}{2}+i \rho} da
\end{equation}
Let $y_{ij}$ be unit tangent vectors, that is, $y_{ij} \in S^{1}$, so that
$\cos(a)= ( \xi_{ij}/ \mid \xi_{ij} \mid ) \cdot y_{ij}$.
As the function $\sinh r/r$
converges to $1$ as $r \rightarrow 0$ we change $\sinh r$ for $r$, which is
just given by $\mid \xi_{ij} \mid$.
Then our kernel can be written as

\begin{equation}
K_{\rho}(x_{i},x_{j})= \int_{S^{1}}
([ x_{i}, x_{j} ] - \xi_{ij} \cdot y_{ij})^{- \frac{1}{2}+i \rho} dy_{ij}
\end{equation}
Our action $S$ is given by

\begin{equation}
S= \sum_{i<j} \rho_{ij} \ln ([ x_{i}, x_{j} ] - \xi_{ij} \cdot y_{ij})
\end{equation}
Now if we first vary the $x$'s and then and then make the $x$'s to be
parallel, that is $\xi_{ij} \rightarrow 0$, we obtain the equations

\begin{eqnarray}
- \rho_{12} y_{12} +  \rho_{23} y_{23} + \rho_{24} y_{24}= 0 \nonumber\\
- \rho_{13} y_{13} - \rho_{23} y_{23} + \rho_{34} y_{34}= 0 \nonumber\\
- \rho_{14} y_{14} - \rho_{24} y_{24} - \rho_{34} y_{34}= 0
\end{eqnarray}
If we sum these three equation we get a fourth one given by
\begin{equation}
\rho_{12} y_{12} +  \rho_{13} y_{13} + \rho_{14} y_{14}= 0
\end{equation}
Each of our equations (33) and (34) describe triangle faces, and the vectors
$y_{ij}$ are two dimensional, as they are vectors in $S^{1}$. Then the equations
describe the faces of a degenerate tetrahedron. This means that our tetrahedron
lies completely in a two-dimensional plane. Once the lengths of edges,
are specified the geometry of the tetrahedron is completely specified.
Therefore
the above equations are not satisfied generically, which implies
that we cannot expect a contribution from degenerate solutions of our
tetrahedron evaluation. This is analogous to the Ponzano-Regge result, in which
they conclude that the only contribution to the 6j-symbol comes from
non-degenerate solutions.
We have a similar argument for our Euclidean case of section 4, so that once
more we agree with Ponzano-Regge.

\danger{Conclusion}. As a conlusion to this entire section
we conclude that the only contribution we get is from non-degenerate
tetrahedra which in accordance to our asymptotic formula (32), and to the
result of the asymptotics of the squared 6j-symbol of \cite{fk} we have
the asymptotic formula

\begin{equation}
T_{6} \sim \frac{1}{\sqrt{V}} \hspace{0.2cm}
\cos \biggl( \sum_{k<l} \rho_{ij} r_{ij}+
k \frac{\pi}{4} \biggr)
\end{equation}
where $V$ is the volume of our tetrahedron.

\section{\bf{Euclidean spin networks}}

With the same technique that we developed for the Lorentzian case,
we can deal with the Euclidean spin networks.
Equation (3) is the partition function for (2+1)-dimensional Euclidean quantum
gravity as a spin foam model, and it is equivalent to the Ponzano-Regge model.

\subsection{SO(3) and its unitary representations}

We now consider the 3-dimensional Euclidean space-time, which is
$\mathbb{R}^{3}$ with the usual positive definite metric given by

\[ [x,y]= x_{0}y_{0} + x_{1}y_{1} + x_{2}y_{2} \]
Consider the two dimensional sphere $S^{2}$, which is the subset of
$\mathbb{R}^{3}$ given by the set of points for which $[x,x]=1$.

$SO(3)$ is the group of unimodular linear transformations which preserve the
positive definite metric. It is compact and connected. The stabilizer of the
unit vector $(1,0,0)$ is isomorphic to $SO(2)$. Therefore our unit sphere
$S^{2}$ can be seen as the homogeneous space $SO(3)/SO(2)$.

The representations of $SO(3)$ are realised as follows. Introduce
spherical coordinates on the unit sphere $S^{2}$. That is, for
$x=(x_{0},x_{1},x_{2}) \in S^{2}$ its coordinates are given by

\begin{eqnarray}
x_{0}= \cos \theta \nonumber \\
x_{1}= \sin \theta \sin \varphi \nonumber \\
x_{2}= \sin \theta \cos \varphi
\end{eqnarray}
In these coordinates the Euclidean Laplace operator

\begin{equation}
\Delta = \frac{\partial^{2}}{\partial x_{0}^{2}}+
\frac{\partial^{2}}{\partial x_{1}^{2}} +
\frac{\partial^{2}}{\partial x_{2}^{2}}
\end{equation}
when restricted to the unit sphere $S^{2}$ is given by

\begin{equation}
\Delta_{0} = (\sinh \theta)^{-1} \frac{\partial}{\partial \theta}
\sin \theta \frac{\partial}{\partial \theta} + (\sinh r)^{-2}
\frac{\partial^{2}}{\partial \varphi^{2}}
\end{equation}
Consider the space of functions on $\mathbb{R}^{3}$.
The group $SO(3)$
acts on this space and the operator of equation (40) commutes with this action.
Then if there is a function $f$ for which $\Delta f(x)=0$ then
$\Delta f(g^{-1} x)=0$, for $g \in S(3)$. A function which satisfies
$\Delta f(x)=0$ is called harmonic. Consider the space of homogeneous
harmonic polynomials
of degree $\ell$ in the variables $x_{0}, x_{1}, x_{2}$. The equality
$T^{\ell}(g)f(x)=f(g^{-1}x)$ defines a representation of $SO(3)$
in the space of homogeneous harmonic polynomials.
These are unitary representations and are labelled by positive integers
$\ell =0,1,2,...$.

\subsection{The kernel}

The question now is, how does the kernel look like? We recall that our
Lorentzian kernel is given by a zonal spherical function of the corresponding
group $SO_{0}(2,1)$, more specifically a Legendre function
$\mathfrak{B}_{-1/2 + i \rho}(\cosh r)$.

It is then obvious to think that for the Euclidean case the kernel is given
by a zonal spherical function of $SO(3)$.
This kernel is then given by the integral

\begin{equation}
K_{\ell}(x,y)= \frac{1}{2 \pi} \int_{0}^{2 \pi}
( \cos \theta + i \cos \varphi \sin \theta)^{\ell} d \varphi
\end{equation}
where $x, y \in S^{2}$.

The above kernel is the well known Legendre polynomial $P_{\ell}(\cos \theta)$
where now $\ell$ denotes an irreducible representation of $SO(3)$.

It happens that the kernel (42) has the following asymptotics

\begin{equation}
K_{\ell}(x,y) \sim \biggl( \frac{2}{\pi \ell \sin \theta} \biggr)^{1/2}
\cos \biggl( (\ell+\frac{1}{2})\theta - \frac{\pi}{4} \biggr)
\end{equation}

\subsection{The tetrahedron network}

We define the evaluation of Euclidean spin networks in the same way we did
it for the Lorentzian case but now using our kernel given by the
the Legendre polynomial $P_{\ell}(\cos \theta)$.
We can also study the asymptotics of the tetrahedron network for this case.
The way to proceed is as we did it for the Lorentzian and Newtonian case, the
only difference being that we now have discrete irreducible representations
$\ell$.

Given our tetrahedron network labelled by unitary irreducible representations of
$SO(3)$, as in figure 2.

\begin{figure}[h]
\begin{center}
\includegraphics[width=0.3\textwidth]{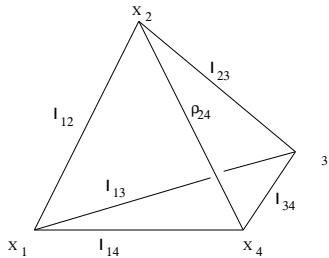}
\caption{Euclidean tetrahedron network}
\end{center}
\end{figure}
we have its evaluation given by the integral (4).

We can as well divide the study of our solutions in non-degenerate and
degenerate ones. According to Ponzano-Regge \cite{pr}, there is no contribution
from degenerate configurations for this Euclidean case. We should be able to
visualise this with the method we follow.

\subsubsection{Non-degenerate solutions}

The tetrahedron network now can be defined by the following integral

\begin{equation}
T_{6}= \int_{({S^{2}})^{4}} \prod_{i<j}
\int_{[0, 2 \pi]^6} e^{ \alpha \sum_{i<j} \ell_{ij}
\ln ( \cos \theta_{ij} + i \cos(\varphi_{ij}) \sinh \theta_{ij})}
d \varphi_{ij} dx_{j}
\end{equation}
Notice that in this case we have an integration carried out over
4 copies of the sphere $S^{2}$, one for each point of our tetrahedron.
This is done because for this case we have a
compact space $S^{2}$, and then there is no divergence when integrating
over these 4 copies.
There is also a very slightly difference here that is interesting to notice.
Our Euclidean tetrahedron integral has a complex integrand.
From first instance it appears that the asymptotics of our integral has to
be treated by the steepest descent method. However if we take that
our action is given by $S= \sum_{i<j} \ell_{ij}
\ln ( \cos \theta_{ij} + i \cos(\varphi_{ij}) \sinh \theta_{ij})$.
we first of all notice that if we vary the action with respect
to the angles $\varphi$'s we find that

\begin{equation}
\frac{\partial S}{\partial \varphi_{ij}}=
\frac{-i \ell_{ij} \sin \varphi_{ij} \sin \theta_{ij}}
{\cos \theta_{ij} +i \cos \varphi_{ij} \sin \theta_{ij}}
\end{equation}
Its stationary points are given by $\varphi_{ij}=0$ and $\varphi_{ij}=\pi$.
Introducing a variable $\epsilon_{ij}=1$ if $\varphi_{ij}=0$, or
$\epsilon_{ij}=-1$ if $\varphi_{ij}=\pi$,
our action can now be written as $S= i \sum_{i<j}
\epsilon_{ij} \ell_{ij} \theta_{ij}$.
Then as an $i$ term has appeared before our action,
it means that now we can deal with
the integral (44) by using the usual stationary phase method.\footnote{We will
notice that this cannot be done when dealing with 'degenerate
configurations'.}

Then, we can vary our action with respect to the four $x$'s variables, paying
attention to the Lagrange multiplier obtaining the following equations

\begin{eqnarray}
\frac{\epsilon_{12} \ell_{12} x_{2}}{\sin \theta_{12}} +
\frac{\epsilon_{13} \ell_{13} x_{3}}{\sin \theta_{13}} +
\frac{\epsilon_{14} \ell_{14} x_{4}}{\sin \theta_{14}}=
\lambda_{1} x_{1} \nonumber\\
\frac{\epsilon_{12} \ell_{12} x_{2}}{\sin \theta_{12}} +
\frac{\epsilon_{23} \ell_{23} x_{3}}{\sin \theta_{23}} +
\frac{\epsilon_{24} \ell_{24} x_{4}}{\sin \theta_{24}}=
\lambda_{2} x_{2} \nonumber\\
\frac{\epsilon_{13} \ell_{13} x_{1}}{\sin \theta_{13}} +
\frac{\epsilon_{23} \ell_{23} x_{2}}{\sin \theta_{23}} +
\frac{\epsilon_{34} \ell_{34} x_{4}}{\sin \theta_{34}}=
\lambda_{3} x_{3} \nonumber\\
\frac{\epsilon_{14} \ell_{14} x_{1}}{\sin \theta_{14}} +
\frac{\epsilon_{24} \ell_{24} x_{2}}{\sin \theta_{24}} +
\frac{\epsilon_{34} \ell_{34} x_{3}}{\sin \theta_{34}}=
\lambda_{4} x_{4}
\end{eqnarray}
We can then take the Euclidean wedge product with $x_{i}$ to the $i$ equation
(i=1,..4), where we take a unit vector $v_{ij}$ in the direction
$x_{i} \wedge x_{j}$. Then we get the following equations

\begin{eqnarray}
\epsilon_{12} \ell_{12} v_{12} +
\epsilon_{13} \ell_{13} v_{13} +
\epsilon_{14} \ell_{14} v_{14}= 0 \nonumber\\
- \epsilon_{12} \ell_{12} v_{12} +
\epsilon_{23} \ell_{23} v_{23} +
\epsilon_{24} \ell_{24} v_{24}= 0 \nonumber\\
- \epsilon_{13} \ell_{13} v_{13} -
\epsilon_{23} \ell_{23} v_{23} +
\epsilon_{34} \ell_{34} v_{34}= 0 \nonumber\\
- \epsilon_{14} \ell_{14} v_{14} -
\epsilon_{24} \ell_{24} v_{24} -
\epsilon_{34} \ell_{34} v_{34}= 0
\end{eqnarray}
If we call $V_{ij}= \epsilon_{ij} \rho_{ij}v_{ij}$, then
the equations (47) imply that the $V_{ij}$ are the edge vectors of
a tetrahedron with edge lengths $\ell_{ij}$.
Each equation describes a face of a
tetrahedron.
Each of our $x_{i}$ normal vectors could be pointing inwards or
outwards. Let $n_{i}$ be the outward normal
pointing vectors to the faces of the tetrahedron. Then we have that
$x_{i}= a_{i} n_{i}$, where $a_{i}= \pm 1$. Recall that our point
$x_{1}$ is fixed at $(1,0,0)$.

All the possible solution configurations are being classified in
\cite{bs}. Moreover, all of our representations are labelled by integers,
which means that the sum of the labels at each vertex is an integer, meaning
that all stationary points either give the same contribution or
give the complex conjugate one. Then given a solution
the corresponding asymptotics is given by

\begin{equation}
T_{6} \sim A \hspace{0.2cm}
\cos \biggl( \sum_{k<l} (\ell_{ij}+ \frac{1}{2}) \theta_{ij}+
k \frac{\pi}{4} \biggr)
\end{equation}
where again $A$ contains information about the determinant of the Hessian, and
k is the signature of it at a stationary point.

The information contained in the factor $A$ must again be related to the volume
of the tetrahedron. This specific calculation can be done numerically for
some examples, but a rigorous proof must be done in a different way.

\danger{Degenerate configurations}.
Degenerate configurations are treated in the same terms we dealt with the
Lorentzian degenerate configurations. All the calculations are similar, and
the conclusion to which we arrive is that
no contribution
is really expected from these configurations.

\section{\bf{Newtonian spin networks}}

Following a similar fashion to our previous Lorentzian and Euclidean case,
we deal now with a case which we call Newtonian. This name is suggested
from the certain limits which will be understood in the present section.

First of all we can just make notice that our Lorentzian and Euclidean kernels
have a similar asymptotics to the $0th$ order Bessel function
$J_{0}(z)$. The $0th$ order Bessel function has an integral
representation given by

\begin{equation}
J_{0}(z)= \frac{1}{2 \pi}\int_{0}^{2 \pi} e^{iz \cos(\theta)} d \theta
\end{equation}
which asymptotics is given by

\begin{equation}
J_{0}(z) \sim \biggl( \frac{2}{\pi z} \biggr)^{1/2}
\cos \biggl( z- \frac{\pi}{4} \biggr)
\end{equation}
This $0th$ order Bessel function is a zonal spherical function for
representations of the group $ISO(2)$ which is the group of motions
of the plane, that is, the group of nonhomogeneous linear transformations
of the plane which preserves distances between points and the orientation.

Mathematically $ISO(2)$ can be obtained by a limit procedure from
$SO_{0}(2,1)$ \cite{vk}. From this limit the
principal unitary irreducible representations of $ISO(2)$
are again given by real numbers $\mathbf{R}^{+}$ and then our
previous kernel $K_{\rho}$ is given by the $0th$ order Bessel function
$J_{0}(\rho r)$ where now our $r$ distance between points is given by the
usual flat 2-dimensional metric on a plane.

Physically this limit corresponds to taking the limit in which the speed of
light goes to infinity ($c \rightarrow \infty)$ .
Then it is natural to think of our Bessel function as
a kernel, which gives us a way to compute spin networks for a Newtonian
quantum gravity model.

\subsection{ISO(2) and its irreducible representations}

Physically we can understand the Newtonian model by consider again Minkowski
space-time $\mathbb{R}^{3}$ with Lorentzian metric, and then taking the limit in which
the speed of light goes to infinity. Then our metric turns into a degenerate
metric with signature $(0,+,+)$. From this limit our hyperbolic space
$H^{2}$ becomes a flat plane isomorphic to $\mathbb{R}^{2}$, and our Lorentz group
$SO(2,1)$ become the group $ISO(2)$.

$ISO(2)$ is the group of motions of the plane $\mathbb{R}^{2}$.
It is the group of nonhomogeneous linear transformations of the plane,
and preserves distances between points, as well as the orientation of the plane.
Every element of $ISO(2)$ is a composition of a rotation about a point
generated by an element of the group $SO(2)$,
and a translation along a vector.
We have that our flat plane $\mathbb{R}^{2}$ is seen as the homogeneous space
$ISO(2)/SO(2)$.

The Laplace operator in $\mathbb{R}^{2}$ is given by

\begin{equation}
\Delta = \frac{\partial^{2}}{\partial x_{1}^{2}}
+ \frac{\partial^{2}}{\partial x_{2}^{2}}
\end{equation}
which in spherical coordinates is given by

\begin{equation}
\Delta = \frac{\partial^{2}}{\partial r^{2}}
+ \frac{1}{r} \frac{\partial}{\partial r}
+ \frac{1}{r^{2}} \frac{\partial^{2}}{\partial \theta^{2}}
\end{equation}
The irreducible representations of $ISO(2)$ can be realised by considering the
space of functions on $\mathbb{R}^{2}$ satisfying the equation
$\Delta f(x)= - R^{2} f(x)$. The equation

\begin{equation}
(T^{\rho}(g)f)(x)=f(g^{-1}x)
\end{equation}
for $\rho= i R$, defines a representation of the group $ISO(2)$ in the space
of functions on $\mathbb{R}^{2}$, which is irreducible for $R \neq 0$, and
unitary when $R \in \mathbb{R}$.

Moreover, the Bessel functions are eigenfunctions of the Laplace operator
on the plane $\mathbb{R}^{2}$.

\subsection{Spin network evaluations}

Before we study the tetrahedron network we give some examples of the evaluation
of spin networks in the Newtonian quantum gravity model.

A Newtonian quantum gravity model in 2+1-dimensions evaluates amplitudes of spin networks
in the 2-dimensional plane $\mathbb{R}^{2}$, and the evaluation of these are given by using
a propagator in this plane which is given by the 0-th order
Bessel function. We follow the usual recipe for evaluating the amplitudes of spin network
graphs which for
our case is given by

\begin{eqnarray}
A(graph)= \prod_{v-1} \int_{(\mathbf{R}^{2})}  dx_{v} \prod_{e}
J_{0}(\rho_{e} r(x,y))
\end{eqnarray}
$\rho_{e}$ denotes a continuous representation of the group $ISO(2)$ which labels an edge,
and $r(x,y)$ is the usual distance between points $x$ and
$y$ in a two dimensional plane.

Examples of the evaluations are given by

\[ \epsfcenter{lor3.eps}= 1 \]

\[ \epsfcenter{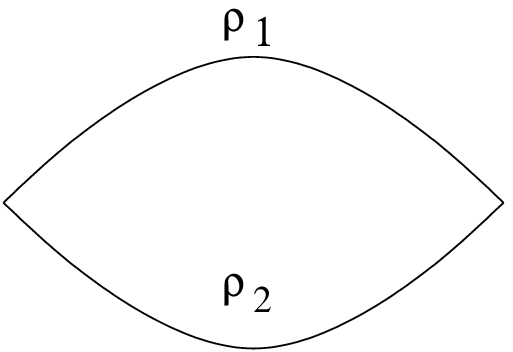}= \frac{\delta(\rho_{2}-\rho_{1})}{\rho_{1}} \]

\[ \epsfcenter{lor2.eps}= \frac{1}{\triangle} \]
where $\triangle$ is the area of the triangle formed by edges of length side
$\rho_{1}, \rho_{2}, \rho_{3}$. If not triangle is formed, the evaluation of the
theta graph is zero.
This shows once more that a convergence of the Lorentzian tetrahedron network
is likely to be expected, since Lorentzian quantum gravity is related to the Newtonian
case, and we can see that for the Newtonian quantum gravity model the
exact evaluation of the theta graph exists.

\subsection{The tetrahedron network}

We can
now define a way to compute a tetrahedron network for this Newtonian case, and
study its asymptotics as we did with the Lorentzian case.

Moreover we can attempt to define a spin foam model of Newtonian quantum
gravity accordingly to the weight amplitudes given to our vertex, edge and face
spin networks.
Particularly we have that a Newtonian tetrahedron network has an evaluation
given by the integral

\begin{eqnarray}
T_{6}= \int_{(\mathbf{R}^{2})^3} dx_{2} dx_{3} dx_{4}
J_{0}(\rho_{12} r(x_{1},x_{2})) J_{0}(\rho_{13} r(x_{1},x_{3}))
J_{0}(\rho_{14} r(x_{1},x_{4})) \nonumber\\
\qquad \qquad \times J_{0}(\rho_{23} r(x_{2},x_{3}))
J_{0}(\rho_{24} r(x_{2},x_{4})) J_{0}(\rho_{34} r(x_{3},x_{4}))
\end{eqnarray}
$\rho_{ij}$ denotes again a continuous representation of the group $ISO(2)$,
and $r(x_{i},x_{j})$ is the euclidean distance between points $x_{i}$ and
$x_{j}$ in a two dimensional plane.
We simplify our notation by denoting $r_{ij}$ for the distance between our
points $x_{i}, x_{j}$.

We can then substitute our integral representations of our Bessel functions
and get

\begin{equation}
T_{6}= \int_{(\mathbf{R}^{2})^{3}} \prod_{i<j}
\int_{[0, 2 \pi]^6} e^{i \sum_{i,j} \rho_{ij} r_{ij} \cos(\theta_{ij})}
d \theta_{ij} dx_{j}
\end{equation}
For this case our points $x_{j}$, for $j=1,,4$ now live on an
embedded flat 2-dimensional plane at our 3-dimensional space-time
at time $1$, where our metric turned into a degenerate one but when
restricted to our 2-dimensional plane, it is just the usual flat Riemannian
metric on that plane.

We proceed similarly as in the Lorentzian and Euclidean cases.

\subsubsection{Non-degenerate solutions}

The exponent $S= \sum_{i<j} \rho_{ij} r_{ij} \cos(\theta_{ij})$ in the
equation above is called the action from which we have to compute the stationary
points in order to obtain the asymptotics of our
Newtonian tetrahedron network.

It is again a similar procedure as the one we followed for the Lorentzian and Euclidean
case, where we find the asymptotic points by varying our action with respect
to the variables $x_{j}$, and $\theta_{ij}$.
For the $\theta$ variables we
just make the variation to be zero.
As for the $x$ variables, we can think of them as vectors with coordinates
$x=(1,X)$, where $X$ is a vector in the $\mathbb{R}^{2}$ plane. The action
can be considered as a function of the $\theta$ variables and of
these two dimensional vectors $X$.

We now look at the stationary points.

The equations we obtain are analogous
to the ones we had for the Lorentzian case and are given by a much simpler
formula

\begin{equation}
\frac{\partial S}{\partial \theta_{ij}}= -\sin \theta_{ij}
\end{equation}
which stationary points are given by $\theta_{ij}=0$ and $\theta_{ij}=\pi$.

Then our action can be written as
$S= \sum_{i<j} \epsilon_{ij} \rho_{ij} r_{ij}$ where $r_{ij}$ is given
just in terms of
the distance in the plane $\mathbb{R}^{2}$ by
$r_{ij}=((X_{i}-X_{j}) \cdot (X_{i}-X_{j}))^{1/2}$.

Varying our action with respect to the variables $X_{2}, X_{3}, X_{4}$,
gives the equations

\begin{eqnarray}
- \frac{\epsilon_{12} \rho_{12} (X_{1}-X_{2})}{r_{12}} +
\frac{\epsilon_{23} \rho_{23} (X_{2}-X_{3})}{r_{23}} +
\frac{\epsilon_{24} \rho_{24} (X_{2}-X_{4})}{r_{24}}=
0 \nonumber\\
- \frac{\epsilon_{13} \rho_{13} (X_{1}-X_{3})}{r_{13}} -
\frac{\epsilon_{23} \rho_{23} (X_{2}-X_{3})}{r_{23}} +
\frac{\epsilon_{34} \rho_{34} (X_{3}-X_{4})}{r_{34}}=
0 \nonumber\\
\frac{\epsilon_{14} \rho_{14} (X_{1}-X_{4})}{r_{14}} +
\frac{\epsilon_{24} \rho_{24} (X_{2}-X_{4})}{r_{24}} +
\frac{\epsilon_{34} \rho_{34} (X_{3}-X_{4})}{r_{34}}=
0
\end{eqnarray}
We note that a solution to the equations (58) is given by the appropiate signs
given to the $\epsilon$'s. Given such solution, interchanging all the signs of
the $\epsilon$'s gives the complex contribution. Then, we can expect an
oscilatory contribution in the cases for which there is a
solution to the equations (58). Moreover, as in this case we have a degenerate
metric, a degenerate tetrahedron is already implicit in the theory.
A better understanding of all the relations between the Netonian theory and
Lorentzian or Euclidean quantum theory is an interesting problem to work in.
This also applies to other dimensions as mentioned already in \cite{bce} for
the 4-dimensional case.

\section{\bf{Conclusions}}

We have observed that all of our kernel functions of the Lorentzian,
Euclidean and Newtonian models are given by
zonal spherical function of a representation of the respective group.
In particular we think
of the Newtonian case as the limit case of Lorentzian quantum gravity
in which the speed of light tends to infinity.
This gives the idea that we have a picture
of a unified theory of quantum gravity and spin networks.

We could generalise all the ideas of the paper in order to deal with
spin networks in any dimension.
We therefore have that Lorentzian spin networks
for n-dimensional quantum gravity whose group is $SO_{0}(n-1 ,1)$ should
be evaluated by using a kernel given by a zonal spherical function of this
group. That is,

\begin{equation}
K_{\rho}^{L}(r)= \frac{\Gamma(\frac{n-1}{2})}
{\sqrt{\pi}\Gamma(\frac{n-2}{2})}
\int_{0}^{\pi}
( \cosh r - \cos \theta \sinh r)^{\sigma} \sin^{n-3} \theta d\theta
\end{equation}
where $\sigma= -p+i \rho$.
$p$ is related to the dimension of the space-time as $p=(n-2)/2$.

Euclidean spin networks for n-dimensional quantum gravity whose group is
$SO(n)$, should then be evaluated by using a zonal spherical function of this
group. That is,

\begin{equation}
K_{\ell}^{E}(\theta)= \frac{\Gamma(\frac{n-1}{2})}
{\sqrt{\pi}\Gamma(\frac{n-2}{2})}
\int_{0}^{\pi}
( \cos \theta - i \cos \varphi \sin \theta)^{\ell} \sin^{n-3} \varphi d\varphi
\end{equation}
And finally Newtonian spin networks for n-dimensional quantum gravity whose group
is $ISO(n)$, should then be evaluated by a zonal spherical function of this
group. That is,

\begin{equation}
K_{\rho}^{N}(r)= \frac{\Gamma(\frac{n-1}{2})}
{\sqrt{\pi}\Gamma(\frac{n-2}{2})}
\int_{0}^{\pi}
e^{r \rho \cos \theta} \sin^{n-3} \theta d\theta
\end{equation}
The above functions are Legendre $\mathfrak{B}_{\sigma}^{n}(\cosh r)$ functions,
Legendre polynomials $P_{\ell}^{n}(\cos \theta)$, and Bessel functions
$J_{n}(\rho r)$ respectively, of the corresponding dimensions.

We can see that for 3-dimensions we have again the kernels we
studied in this paper.

For the 4-dimensional case, we have $\sigma =-1+i \rho$, that is, a
principal unitary series representation of the Lorentz group $SO_{0}(3,1)$,
then our Lorentzian kernel reduces to

\begin{equation}
K_{\rho}^{L}(r)= \frac{\sin(\rho \theta)}{\rho \sinh \theta}
\end{equation}
which is the Barrett-Crane one.

Similarly for the Euclidean and Newtonian kernels, we get

\begin{equation}
K_{\rho}^{E}(\theta)= \frac{\sin(\ell +1) \varphi}{(\ell +1) \sin \varphi}
\end{equation}

\begin{equation}
K_{\rho}^{N}(r)= \frac{\sin(\rho r)}{\rho r}
\end{equation}
The way all these kernels are related is given in \cite{vk}. In particular
for our 3-dimensional case we have that

\begin{equation}
\lim_{t \rightarrow \infty}
K_{\rho}^{L}(r/t) = J_{0}(r)
\end{equation}

\begin{equation}
\lim_{\ell \rightarrow \infty}
K_{\ell}^{E}(\theta / \ell) =J_{0}(r)
\end{equation}

This is interpreted as saying that when dealing with the degenerate configurations
in the Lorentzian and Euclidean spin network evaluations, these can be
related to the Newtonian case.
A better understanding of what all this implies to the models is an interesting
problem. In particular what we can now about the physics of these models is important.
We want to stress once more the interest in the Newtonian model. In the end, it is
related to the Lorentzian and Euclidean, and it is simpler.
It is then interesting to study Newtonian quantum gravity as itself and try to
find what it could be telling us about the physics of quantum gravity.

\vspace{1cm}

\danger{Acknowledgements}

I want to thank to my supervisor John Barrett for his time on many helpful discussions.
Also I want to thank Laurent Freidel and Kirill Krasnov for
discussing the convergence problem of the tetrahedron evaluation.

This work was supported in part by CONACYT grant 132375, by an ORS award and
by a full tuition scholarship from the University of Nottingham.

\end{document}